# Towards Security of Additive Layer Manufacturing

Mark Yampolskiy, Todd R. Andel, J. Todd McDonald, William B. Glisson, Alec Yasinsac
University of South Alabama

## ABSTRACT

Additive Layer Manufacturing (ALM), also broadly known as 3D printing, is a new technology to produce 3D objects. As an opposite approach to the conventional subtractive manufacturing process, 3D objects are created by adding thin material layers over layers. Until recently, they have been used, mainly, for plastic models. However, the technology has evolved making it possible to use high-quality printing with metal alloys. Agencies and companies like NASA, ESA, Boeing, Airbus, etc. are investigating various ALM technology application areas. Recently, SpaceX used additive manufacturing to produce engine chambers for the newest Dragon spacecraft. BAE System plans to print on-demand a complete Unmanned Aerial Vehicle (UAV), depending on the operational requirements.

Companies expect the implementation of ALM technology will bring a broad variety of technological and economic benefits. This includes, but not limited to, the reduction of the time needed to produce complex parts, reduction of wasted material and thus control of production costs along with minimization of part storage space as companies implement just-in-time and on-demand production solutions. The broad variety of application areas and a high grade of computerization of the manufacturing process will inevitably make ALM an attractive target for various attacks.

**Compliance to Customer Specification**

Currently, three main ALM techniques are used to work with metals and alloys: powder bed fusion process, direct metal deposition, and metal sheet lamination. They differ in source material (e.g., wire, powder, and sheets), principles of source material distribution (e.g., nozzle, powder bed) and heat source (e.g., electronic beam, laser). Depending on the chosen ALM technique, different manufacturing process parameters can influence the quality of a part that has been manufactured. Malicious manipulation of the manufacturing parameters can be viewed as the most dangerous kind of attack on ALM. The reason is that such attacks can alter the material properties of the manufactured object (e.g., its resistance to mechanical stress) without altering the 3D shape. If such an object is used, in safety critical devices like rocket engines or airplane turbines, such degradations can easily lead to loss of life and negatively impact business revenues.

Our extensive survey of ALM related literature reveals numerous vulnerabilities in ALM equipment [1]. The analysis identifies which attack vectors can be exploited for various influences on manufacturing processes, thus causing consequential impacts on the manufactured 3D objects. This, in turn, can be exploited as a basis for adversarial business models. This vulnerability analysis provides the groundwork for ongoing and future research.

We are investigating three research directions. First, we are examining preparation for experimental penetration testing, which should (i) evaluate the feasibility of the identified attack and (ii) assess the difficulty thereof. Second, we plan to evaluate the achievable effect of particular attacks on the printed 3D object, thus providing the basis for the assessment of the possible business impact. Last, but not least, we will develop concepts and strategies aimed at preventing, detecting, and – if the first two have failed – reconstructing successful attacks through examination of residual data and digital forensic analysis.

**Intellectual Property Protection**

Orthogonal to the customer's specification compliance is the problem of the Intellectual Property Protection. 3D objects, which are manufactured with ALM technology, are usually integral parts of complex machines, e.g., blades of turbines. As such, both the 3D shape as well as the specification of the required physical properties, e.g., capability to withstand specified mechanical or thermal influences, is the designer's Intellectual Property (IP).

However due to high equipment costs, designers of the 3D objects rarely have their own ALM capabilities and, therefore, order production through third parties specializing in ALM. At this point, tuning of the ALM manufacturing parameters in order to satisfy specific customer requirements might be necessary. The specific combination of the manufacturing parameters is the IP of the company specializing in ALM production. The manufacturing process related information is rarely shared with the customers ordering 3D printed objects.

We have designed a flexible outsourcing model to increase competition as well as collaboration between companies which specialize in different manufacturing aspects. In this model [2], companies specializing in different aspects handle their IP licenses. This model supports a binding between the IP (3D object shape and required properties, and manufacturing parameters), the actors issuing and using licenses for IP (owner of the 3D object design, tuning experts, owner of ALM equipment), and limitations imposed on the licenses (number of 'runs' for which a particular IP could be legally implemented). The article lists numerous requirements which should be addressed by technical means in order to enforce IP protection in the proposed model.

Currently, we are investigating various technical solutions to enforce the proposed secure outsourcing model. This includes specialized cryptographic protocols, obfuscation of hardware (in order to prevent reverse engineering), as well as non-invasive monitoring and tamper-resistant IP recording techniques to protect relevant information, which should enable detection of IP violations. This environment encourages future research into effective digital forensics methodologies and solutions in ALM environments.